\documentstyle[12pt,aasms4]{article}

\newcommand{\ea}{{\it et al.~}}

\lefthead{Vilhu et al.}
\righthead{UV Spectroscopy of AB Doradus}

\begin{document}

\title{ UV Spectroscopy of AB Doradus with the  Hubble Space 
Telescope.\footnotemark[1] \\
Impulsive flares and bimodal profiles of \\ 
	     the CIV 1549  line in a young star}

\author{O. Vilhu\altaffilmark{2}}
\affil{NORDITA, Blegdamsvej 17, DK-2100 Copenhagen, Denmark, and Observatory, 
   Box 14, FIN-00014 University of Helsinki, Finland\\
      Electronic mail: osmi.vilhu@helsinki.fi}
\author{P. Muhli,  J. Huovelin and P. Hakala}
\affil{Observatory, P.O. Box 14, FIN-00014 University of Helsinki, Finland\\
	Electronic mail: 
	muhli@gstar.astro.helsinki.fi, 
huovelin@fornax.astro.helsinki.fi, pahakala@sirius.astro.helsinki.fi}

\author{S.M. Rucinski}
\affil{David Dunlap Observatory, University of Toronto, P.O. Box 360
Richmond Hill, Ontario, Canada LYC 4YC\\
       Electronic mail: rucinski@astro.utoronto.ca}

\and

\author{A. Collier Cameron}
\affil{School of Physics and Astronomy, University of St Andrews,
North Haugh, St Andrews, Five, Scotland KY16 9SS\\
       Electronic mail: acc4@st-andrews.ac.uk}

\footnotetext[1]{Based on observations with the NASA/ESA {\it Hubble Space Telescope},
obtained at the Space Telescope Science Institute which is operated
by the Association of Universities for Research in Astronomy, Inc., under NASA 
Contract NAS5-26555.}

\altaffiltext{2}{{\it Hubble Space Telescope} Guest Observer}

\begin{abstract} 
We observed AB Doradus, a young and active late type star
(K0 - K2 IV-V, P = 0.514 d) with the Goddard High Resolution Spectrograph of the
post-COSTAR {\it Hubble Space Telescope}
with the time and spectral resolutions of 27 s and 15 km
s$^{-1}$, respectively (November 14.08 - 14.30, 1994 (UT)).
The wavelength band  (1531 - 1565 \AA) included the strong 
CIV doublet (1548.202 and 1550.774 \AA, formed in the transition
region at 10$^5$ K), the  chromospheric SiII 1533.432 \AA\ 
line and the blend of SiI, CI and FeII lines at 1561 \AA.
The mean quiescent CIV flux state was characterized by
F$_{CIV}$ =  (7.80 $\pm{0.34}$) $\times$ 10$^5$ ergs cm$^{-2}$
s$^{-1}$, close to the saturated value and 100 times the solar one. 
The  line profile (after removing the rotational and instrumental profiles) 
is bimodal consisting of two Gaussians, narrow 
(FWHM = 70 km s$^{-1}$)  and broad (FWHM = 330 km s$^{-1}$). This
bimodality is probably due to two separate broadening 
mechanisms and  velocity fields at the coronal base. 
It is possible that TR transient events (random multiple velocities), with a large 
surface coverage, give rise to the broadening of the narrow component,
while true microflaring is responsible for the broad one as suggested by
Wood, Linsky and Ayres (1997).

The transition region was observed to
flare  frequently on different time scales and 
magnitudes.
The largest impulsive flare seen in the CIV 1549 \AA\ emission 
at day 14.22 reached in less than one minute the peak
differential emission measure N$_{e}^2$V (10$^{4.85}$ K -- 10$^{5.15}$
K) = 10$^{51.2}$ cm$^{-3}$ and returned exponentially in 5 minutes 
to the 7 times lower 
quiescent level.
The 3 min average line profile of the flare was
blue-shifted (--190 km s$^{-1}$) and broadened   
(FWHM = 800 km s$^{-1}$). This impulsive flare could have been due to 
a chromospheric heating and subsequent evaporation 
by an electron beam, accelerated (by reconnection) at the apex of a 
coronal loop. 
\end{abstract}
   

\section{Introduction}

AB Doradus (HD 36705) is one of the most interesting
active late-type stars observed from radio to X-rays. 
It has  short rotation period (0.514 d), small age 10$^{7-8}$ yr
and spectral type K0-K2 IV-V (for some recent works see e.g. \markcite{vilh93}
Vilhu \ea 1993; \markcite{ruci95} Rucinski \ea 1995; \markcite{whit95} White \ea 1996).
The new distance 15 pc is based on the combined HIPPARCOS data and
VLBI radio techniques giving a parallax
0.0663-0.0672 arc sec (Guirado et al. 1997). These astrometric observations also
discovered another low mass companion (around 0.1 solar masses) of AB Dor
at separation of 0.2-0.7 arc sec,
in addition to the known visual companion Rst137B at 10 arc sec.

Striking features are the frequently observed
X-ray flares (twice/day, see Vilhu et al. 1993), 
and the persistent (over 15 years) photometric 
ephemeris, explained by spots at active longitudes (Innis et al. 1988):

\begin{equation}
\rm{PHASE = 0.0: HJD}=2444296.575+0.51479E  
\label{eq:eph}
\end{equation}

An interesting discovery were the prominence-like
condensations trapped in corotation at  several stellar radii above
the surface \markcite{came89} (Cameron \& Robinson 1989).
  At least a part of the stellar angular momentum loss takes place
through these clouds.

Many studies have demonstrated the usefulness of the strong CIV 1549
\AA~ doublet (formed at the transition region) for studying the magnetic 
activity (dynamo power) in solar-type 
stars. The emission  saturates
at around F$_{CIV}$/F$_{BOL}$ = 2.5 $\times$ 10$^{-5}$  \markcite{vilh87}
(Vilhu 1987). 
One of the major discoveries of the GHRS (Goddard High Resolution Spectrograph)
on board the {\it Hubble Space Telescope} has been the detection of bimodal 
structures and broad wings of the CIV 1549 lines in some active cool stars.
\markcite{lins94} Linsky \& Wood (1994) were the first to report broad
CIV and SiIV line
profiles in the flare star AU Mic (dM0), even during the quiescent state. They
fitted the quiescent  spectrum of AU Mic by two components
with relative fluxes F$_B$/F$_N$ = 0.58 and FWHM$_B$/FWHM$_N$ =
173 km s$^{-1}$/29 km s$^{-1}$.
The quiescent CIV flux was high (6.3 $\times$ 10$^5$ ergs cm$^{-2}$ s$^{-1}$),
somewhat higher than the 'maximum saturated' value at M0 (2 $\times$ 10$^5$ ergs cm$^{-2}$ s$^{-1}$).  

The second example, Capella (a binary star, G1 III + G8 III), was observed by
\markcite{lins95} Linsky \ea (1995). The G1-star (vsin{\it i} =  36
km s$^{-1}$) contributes 88 \% of the CIV flux
from the system, and is composed of a moderately broad (FWHM = 164 km s$^{-1}$) 
and a very broad component (FWHM = 360 km s$^{-1}$), with flux ratios around 0.8.
The mean surface flux of the G1-component was roughly F$_{CIV}$ = 
1.5 $\times$ 10$^5$ ergs cm$^{-2}$ s$^{-1}$, which is 10 times lower than the 
saturated value at this spectral type, but still higher than that of a solar 
plage (10$^{4.5 - 5.0}$ ergs cm$^{-2}$ s$^{-1}$). \markcite{lins95}
Linsky \ea proposed that the very broad wings of the transition 
region lines of Capella are due to stellar analogs of solar transition 
region explosive events (Brueckner at al. 1988, Dere et al. 1989).
\markcite{wood95} Wood \ea (1996) analyzed the GHRS
observations of HR 1099
and found one more example of very broad wings in the transition region lines.
   
Later, the number of stars observed has increased. At present, the most
comprehensive collection of broad wings is in the important paper by 
Wood, Linsky and Ayres (1997), including 12 stars with different activity
levels and spectral and luminosity classes (including the measurements of AB Dor
of the present paper). In particular, it was shown that the contribution of
the broad component correlates with activity indicators like the X-ray surface
flux. 

Preliminary results of the present study 
were reported by \markcite{vilh95} Vilhu \ea (1996).
In the present paper we describe these observations 
of AB Doradus in a more detail.
 
\section{Observations}

The observations (see Table 1) covered November 14.08 - 14.30, 1994 (UT)
corresponding to rotational phases between 0.21 - 0.64 (Eq. \ref{eq:eph}). 
The observations were carried out during the Continuous Viewing Zone (CVZ),
with no iterrupted sequences owing to Earth occultations.
The post-COSTAR GHRS (cycle 4) was used with the medium resolution 
grating G160M, centered
at the CIV 1550 doublet, giving resolving power of 20000 (see GHRS 
Instrument Handbook). 
The ACCUM mode with STEP-PATT = 5 gave time resolution
of 27 s. 

The wavelength coverage was 1531 - 1565 \AA~ with 
a diode spacing in the substep pattern 
of 0.019 \AA~ (3.7 km s$^{-1}$). In addition to
the strong CIV 1548.202 and CIV 1550.774 lines (the optically thin theoretical 
ratio of which equals to 2), the region contains also the well visible 
chromospheric SiII 1533.432 line
and the broad blend at 1560 \AA, consisting of chromospheric 
SiI, CI and FeII lines (see Fig. \ref{fig:averspec}). 

The instrumental profile was Gaussian with a  FWHM approximately 15 km s$^{-1}$. 
The wavelength calibration was based on the exposure of the calibration lamp
just before the observing run.
The magnetic anomalies (GIMP) were estimated (by D. Soderblom, 
from the model hrs-gimp
which uses spacecraft position) and their 
effect on the wavelength scale was found small (less than $\pm$ 0.01 \AA),
although occasionally it could reach a quarter of a diode (0.019 \AA = 3.7 
km/s).
Further, the temperature corrections applied by the calhrs-routine has been
generally good, leading to no significant zero-point nor dispersion error
(D. Soderblom, private comm.)

The strong CIV doublet (quiescent flux $\sim$ 1.0 $\times$ 10$^{-12}$ ergs 
cm$^{-2}$ s$^{-1}$, see Table 2) is 
contaminated by weak chromospheric SiI lines 
(at 1548.978, 1551.240, 1551.454 and
1551.856 \AA). However, their effect is very small and can be neglected
because in active stars the transition region lines are much more enhanced 
than the chromospheric ones. In solar plages, resembling the outer atmospheres 
of moderately active stars, these lines comprise 4 \% of the CIV line fluxes.
In the extremely active AB Dor we can expect this percentage to be still lower.
The weaknesses of the SiII 1533.432 line  (1.8 $\times$ 10$^{-14}$ 
ergs cm$^{-2}$ s$^{-1}$) and the 1560-blend (7.1 $\times$ 10$^{-14}$ ergs 
cm$^{-2}$ s$^{-1}$) also confirm this: in the quiet sun and solar plages these 
lines are stronger than the SiI lines inside the CIV-profile 
(Fig. \ref{fig:averspec}).

\placetable{tbl:1}

\placefigure{fig:averspec}

The single SiII 1533.432 line (see Fig. \ref{fig:averspec}) can be well fitted 
with the 90 km s$^{-1}$
rotational and 15 km s$^{-1}$ instrumental profiles. Hence, these
(previously known) broadening values are also used in the present paper.

A photospheric dark spot (the spot 'B' at phase 0.5, defined by 
Innis et al. 1988) was well visible during 
our observations, causing the minimum in the simultaneous ground-based optical 
U-band light curve (Fig. \ref{fig:Ucurve}, Las Campanas/University 
of Toronto 61 cm telescope). 

\placefigure{fig:Ucurve}

\placefigure{fig:2D}

\placefigure{fig:CIV}

\placetable{tbl:2}

\placetable{tbl:3}

Figures \ref{fig:2D} and \ref{fig:CIV} show the time-evolution of the CIV 
line during the observations (5.4 hours).  
Especially remarkable are the impulsive flares (see the next section).

\section{Impulsive Flaring}

A striking feature of the CIV line is the frequent flaring (Fig. \ref{fig:2D} and 
\ref{fig:CIV}).
The line intensity is variable on different time scales: AB Dor spends 
approximately 15 \% of the time above the  + 3 $\sigma$ level from the mean
quiescent state, defined as the CIV flux-level averaged over the rotational phases
0.25-0.35 and 0.50-0.60 without significant flares.
On the other hand, the flares were not detected in the far UV continuum close 
to  CIV 1550 (these observations) and in the optical U-band continuum (see 
Fig. \ref{fig:Ucurve}). In addition, the flux of the nearby chromospheric lines 
did not increase during the flares at all. These constant lines were
the single SiII 1533.432 
line, and the blend of SiI, CI and FeII lines at 1561 \AA.

The CIV light curves around the two strongest flares at day 14.185 and 14.22
were fitted with an exponential function which has also been used for
modelling of the temporal behaviour of Gamma Ray Bursts (see
 \markcite{norr94} Norris \ea 1994):

\begin{equation}
F(t) = A \, \exp [ - ( \vert t - t_{max} \vert / \sigma_{r,d} )^{\nu}
] + B,
\label{eq:grbfit}
\end{equation}
 
where $A$ denotes the maximum intensity of the flare relative to the
quiescent intensity $B$, and $t_{max}$ is the corresponding time
coordinate. $\sigma_r$ and $\sigma_d$ are the rise ($t < t_{max}$) and
decay ($t > t_{max}$) times of the flare, respectively, and $\nu$ is
the flare ``peakedness''. A least squares fit was applied to
the CIV light curve between day fractions 14.173 - 14.196 and 14.207 -
14.230 (corresponding to rotation phase intervals of 0.394-0.406 and
0.462-0.474, respectively), yielding the
following results for the day 14.185 and 14.22 flares, respectively
(errors are $1 \sigma$): $A = 1.93 \pm 0.17$ and $7.18 \pm 0.93$ ergs
cm$^{-2}$ s$^{-1}$, $B = 1.52 \pm 0.04$ and $1.20 \pm 0.04$ ergs
cm$^{-2}$ s$^{-1}$, $t_{max} = 14.1837 \pm 0.0002$ and $14.2179 \pm
0.0000$ UT, $\sigma_r = 115 \pm 24$ and $26 \pm 4$ seconds, $\sigma_d
= 229 \pm 30$ and $96 \pm 17$ seconds, $\nu = 1.90 \pm 0.42$ and
$0.94 \pm 0.14$. The CIV light curves and the corresponding fits are
shown in Figs. \ref{fig:flarefits}a and \ref{fig:flarefits}b.

\placefigure{fig:flarefits}

The strongest flare (rise time 26 s, decay time 96 s) at day 14.22 is well 
represented (3 min mean)  by the quiescent
profile {\it plus} a very broad  (FWHM = 800 km s$^{-1}$) and
blue-shifted (--190 km s$^{-1}$) Gaussian profile (see
Fig. \ref{fig:lineprofs}~ and Table 3). This impulsive flare deserves a special
discussion. The error estimate of its rise time ($26\pm 4 s$) is probably  
too optimistic and formal given by the fitting method (see Table 2). 
However, the  
rise time can in principle be determided with a better accuracy than the time
resolution (like the wavelength can often
be measured more accurately than  the spectral resolution). 
The result is, however, marked by (:) in the Table 2. 

During the flare (and during the whole observing run) the line ratio 
CIV1548/CIV1551 remained constant (around 1.9), close to the theoretical
non-saturated value 2.0. This indicates that the flaring plasma remained 
effectively thin (the line photons once created escape from the region after 
perhaps many scatterings).
 The narrow (FWHM = 55 km s$^{-1}$) and slightly blue-shifted
(--50 km s$^{-1}$) absorptions seen during the flare (see Fig. \ref{fig:lineprofs}) 
had approximate equivalent widths
0.3 \AA~ and 0.2 \AA~, respectively (after the 
quiescent component was subtracted). This would mean that the flaring upwards 
moving plasma became somewhat optically thick (although remaining effectively 
thin), or alternatively the absorptions were caused by some unlucky
short-term instrumental effect (like intermittent diodes). 
This last possibility is unlikely because it is seen in both line components
(see Fig.6).

A  solar flare with similar time-evolution was observed by \markcite{brek95}
Brekke \ea (1996) with the Solar-Stellar Irradiance Comparison Experiment
(SOLSTICE). The  rise time of that flare was 10 s and it decayed in a few 
minutes. The peak intensity of CIV 1549 emission reached around 12.5 times the 
pre-flare value (Brekke et al., their Table 2).    
In AB Dor the flare peak CIV-flux reached 6.6 times 
its quiet 
value 7.8 $\times$ 10$^5$ ergs cm$^{-2}$ s$^{-1}$.  
Using for the quiet sun a surface flux of 10$^{3.7-4.0}$
ergs cm$^{-2}$ s$^{-1}$, and assuming that the radii of the sun and AB Dor are
comparable, one concludes that the AB Dor flare was approximately
40 - 80 times stronger than the solar SOLSTICE flare. 

\markcite{mari89}
Mariska \ea (1989) have made interesting hydrodynamical computations
of impulsive
flares (see also \markcite{mari85} Mariska \& Poland 1985). Their results are
relevant to explain
the main features of our flare at day 14.22, particularly the short time scale
of the event. The model assumes that the
impulsive phase of the flare is due to the heating and subsequent evaporation
of the stellar surface by an energetic electron beam  accelerated 
(by reconnection) at the apex of a coronal loop. During the beaming
the chromosphere is heated and changed to a high density transition region.
Initially the CIV 1549 line is formed in typical TR-conditions 
(electron densities
10$^{10}$ cm$^{-3}$) but during the flare its formation  density is rising 
to 10$^{12}$ cm$^{-3}$.
 Impulsive high temperature lines
(like CIV) are radiated from this collisionally heated plasma 
together with a hard X-ray pulse (Bremsstrahlung).
The maximum heating occurs around 10$^5$ K, i.e. where the CIV 1549 line 
is formed.

The reference model of \markcite{mari89} Mariska \ea (1989) 
can be scaled to reproduce the main characteristics 
of the day 14.22 flare, although one can not prove that the flare was
exactly similar to the model.  The electron beam flux was 
increased  in the computations
linearly in 30 s to 5 $\times$ 10$^{10}$ ergs cm$^{-2}$ s$^{-1}$ and after 
that
in 30 s decreased to zero again. The hardness of the electron beam was
characterized by the low energy cut-off 15 keV and spectral power law 
index $\delta$ = 6.
However, the results of Mariska et al. were quite unsensitive on the
choise of $\delta$ ($\delta$=4 gave practically the same
results). Most important parameters were the injection flux and
the time scale.

Ten seconds after the beginning of the heating the differential emission measure
at 10$^{5}$ K
increased to 10$^{48}$ cm$^{-3}$, assuming (in the models) a flaring 
loop with 3000 km
cross section (the hydrodynamics itself is completely independent of this
selection). If the cross sectional area would have been  1900 - 3800
 times larger, 
the observed AB Dor flare maximum differential emission measure 
10$^{51.3 - 51.6}$ 
cm$^{-3}$ could be achieved. This means 0.5 - 1.0 \%  flare 
coverage of the total
stellar surface.

Using the \markcite{mari89} Mariska \ea~ electron flux parameters one
can estimate the total
number of injected electrons as  10$^{39.9 - 40.2}$ during the flare at day 
14.22 (by integrating over the electron injection time and energy profiles, 
 eqs. 9 and 10 in \markcite{mari89} Mariska \ea).  
At typical coronal loop electron densities (10$^{10}$ cm$^{-3}$) this 
amount of electrons is already existing in a reservoir 
volume of 10$^{29.9 - 30.2}$
cm$^3$ (before the reconnection). 
For the loop foot point radius estimated above
((0.65 - 0.92) $\times$ 10$^{10}$ cm = (0.09 - 0.12) $\times$ R$_{\sun}$) 
this means a (0.06 - 0.17) $\times$  R$_{\sun}$ long piece of a flux tube
as the reservoir of the injected electrons,  
e.g. at the apex of an already existing coronal loop.

\placefigure{fig:lineprofs}


\section{Bimodal Line Profiles. Narrow and Broad Components.}

The accumulated quiescent profile of the CIV doublet
(with flares extracted and shown in  Fig. \ref{fig:lineprofs}) can be
fitted with four 
Gaussians (at the non-rotating stellar surface), broadened with the
90 km s$^{-1}$ rotational (Gray function) 
and 15 km s$^{-1}$ instrumental (Gaussian)  profiles
(narrow and broad components at each line of the CIV doublet). In this way
we have assumed that the transition region lies close to the photosphere,
which is the case for normal coronal loops.     However, it is 
 possible that AB Dor has a co-rotating
'extended' transition region with a larger rotation velocity which may
 explain
the width of the narrow component.

The linear limb darkening coefficient was assumed to be zero 
(no darkening nor brightening). The results 
were found to be quite insensitive to this choise. Moving   
the coeffiecient between +1 and -1, produced only 5 per cent changes in
the intrinsic FWHM-values. 

The mean fitting parameters (radial velocity relative to the stellar surface, 
FWHM, flux at the stellar surface and differential emission measure)
for the whole doublet are presented in Table 3. 
30 km s$^{-1}$ was used as the radial velocity of the star itself 
(Vilhu \ea 1987).
The narrow and broad components have roughly equal fluxes in AB Dor and
both indicate large non-thermal velocities, much larger than the 15
km s$^{-1}$ thermal velocity of CIV at 10$^5$ K. 

\section{Discussion}

Using low resolution X-ray spectroscopy with the {\it Einstein} 
and {\it Exosat}, coronae of active stars were best fitted by two thermal
components with temperatures around (3-7) $\times$ 10$^6$ and (2-5) $\times$ 
10$^7$ K, respectively, with varying relative emission measures 
(e.g. \markcite{swan81} Swank 
et al 1981; \markcite{schm90} Schmitt \ea 1990; \markcite{leme89} Lemen \ea
1989), although the  reality of  this bimodality is  somewhat in doubt. 
If the coronal gas is fitted with one single (mean) temperature,  
a clear correlation 
between the temperature and activity level exists (Gagne et al. 1995).
In the 2-T models the relative strength of the hot component rises with the
CIV flux intensity \markcite{vilh90} (Vilhu \& Linsky 1990). 
A similar trend was found by \markcite{gagn95}
Gagne \ea  in the Pleiades: the coronal mean temperature clearly
correlated with L$_{X}$/L$_{bol}$.

Using ASCA, 
\markcite{whit95} White \ea (1996) fitted the quiescent spectrum 
of AB Dor with a 2-T plasma (
7 $\times$ 10$^6$ K and
1.7 $\times$ 10$^7$ K) with roughly equal emission measures for both
components. Combining
the EUVE spectrum with that of ASCA \markcite{mewe95} (Mewe \ea 1996)
gives best fits for 7.2 $\times$ 10$^6$ K
and 1.9 $\times$ 10$^7$ K with emission measures 9.0 and 6.5 (in units of
10$^{52}$ cm$^{-3}$), respectively (for the iron abundance 0.32 $\times$ 
solar value). We found  roughly equal emission measures for the 
narrow and broad components of CIV. 
Hence, it is tempting to speculate that the two velocity 
fields we  see in the 
CIV-profiles of AB Dor are related to separate 
velocity field 
regions at the bases of
two coronal patterns, hot and cool. However, multitemperature fits are also possible as demonstrated by
Dupree et al. (1993) for Capella, using EUVE.

No trace of this bimodality or extra non-rotational broadening can be seen
in the nearby chromospheric SiII 1533.4 line, nor in the earlier observations
of photospheric absorption lines. This
supports the idea that the broadening is related to motions in the transition
region only. 
However, the MgII 2800 k and h resonance lines are somewhat 
broader than expected from
the rotational broadening only. Rucinski (1985), using the IUE, found
FWHM = 159$\pm{4}$ km s$^{-1}$ and = 150$\pm{4}$ km s$^{-1}$ for the k and h lines,
respectively. At the line bases the total widths were 289$\pm{9}$ km s$^{-1}$ and
232$\pm{9}$ km s$^{-1}$ corresponding to rotational velocities vsin$\it{i}$ =
142 -- 116 km s$^{-1}$. Since the instrumental profile of the IUE high resolution
spectrograph at 2800 \AA~ was around 20--30 km s$^{-1}$, some non-thermal velocities
might be present but not so large as observed in the CIV 1550 lines.

Linsky et al. (1995) and Wood et al. (1997) proposed that microflaring
(solar explosive events type phenomena) is behind the broad component.
However, there is another possible hypothetical mechanism to explain the broad 
component (observed Gaussian FWHM = 330 km s$^{-1}$)
we see in the quiescent spectrum of Fig. \ref{fig:lineprofs}. Suppose that 
this component is not arising in large X-ray loop structures at all, but 
within the slingshot prominence complexes we see in H$\alpha$, discovered by
\markcite{came89} Collier Cameron \& 
Robinson (1989).
The co-rotating slingshot prominence system we see transiting
 the disc of AB Dor is concentrated around 2.7 to 3 stellar radii from the 
rotational axis
(where the centrifugal force equals the gravity). 
This gives each individual cloud a velocity half-amplitude of 
270 km s$^{-1}$. A large enough number of these could well mimic a Gaussian with 
FWHM in the required range.
However, this mechanism is highly speculative since we have no physical
model to explain why the prominence system should contain 10$^5$ K gas.
Hence, the physics behind the broad component is likely to be real microflaring
(solar explosive events type microflares) as suggested by Linsky and 
collaborators.

To explain the broadening of the narrow component (FWHM = 68$\pm{2}$ km s$^{-1}$,
including the 15 km s$^{-1}$ thermal broadening), 
it is tempting to look also at the Sun. 
Dere \& Mason (1993) fitted the average solar CIV-profile with a non-thermal 
component with FWHM = 27$\pm{5}$ km s$^{-1}$. 
Adding the thermal broadening, a total 
FWHM = 42$\pm{5}$ km s$^{-1}$ results. Alternatively, 
they could fit the profile with two 
Gaussians (FWHM = 33 km s$^{-1}$ and 66 km s$^{-1}$, 
with relative intensities 1:2).

Random multiple velocities as seen in the high spatial and spectral resolution
CIV Spacelab-data by Brynildsen et al (1995) are probably the physical reason for 
this extra non-thermal  broadening of the narrow component.
 We co-added these Spacelab-data 
spatial pixels and the average line profile could be well fitted with a 
Gaussian profile with FWHM = 39$\pm{5}$ km s$^{-1}$
(the Spacelab data by Brynildsen, private comm.), confirming the result by
Dere and Mason. The same 
broadening was obtained for both a quiet and 
an active region, which was somewhat a surprise.

However, this need not to be the final word. One can still speculate
about extended transition regions and slingshot prominences as 
sources for these broadenings, at least in the case of AB Dor.   

\section{Conclusions}

We observed AB Doradus with the GHRS spectrograph of the {\it Hubble Space
Telescope} on November 14.08 - 14.30, 1994 (UT). The medium resolution grating
G160M was used with time and spectral resolution  of 27 s and 15 km s$^{-1}$,
respectively (post-COSTAR).
The wavelength band observed (1531 - 1565 \AA) included the strong 
CIV doublet (1548.202 and 1550.774 \AA, formed in the transition region at 
10$^5$ K), the weak chromospheric SiII 1533.432 line and a blend of SiI, CI 
and FeII lines at 1561 \AA. The following main results were obtained:
   
1. The quiescent state of AB Dor (see Table 3) is characterized by the CIV 
surface flux
F$_{CIV}$ =  (7.80 $\pm{0.34}$) $\times$ 10$^5$ ergs cm$^{-2}$ s$^{-1}$. 
This is almost the same as the ``saturated'' value at K0-2 
spectral type (8 $\times$ 10$^5$ 
ergs cm$^{-2}$ s$^{-1}$, see \markcite{vilh87} Vilhu 1987).
 The quiescent CIV 1549 line profile of AB Dor is bimodal and was 
fitted with two Gaussian profiles (narrow and broad),
after the rotational and instrumental profiles were removed. 
The best fits were (for the line centroid radial velocity ${\it}$
minus the stellar radial velocity 30 km s$^{-1}$, FWHM and total surface flux): NARROW
(8 $\pm{1}$ km s$^{-1}$, 68 $\pm{2}$ km s$^{-1}$, 4.00 $\pm{0.16}$ $\times$ 10$^{5}$ 
$\rm{ergs} ~\rm{cm}^{-2} ~\rm{s}^{-1}$) and  BROAD (2 $\pm{4}$ km s$^{-1}$, 334 
$\pm{8}$ km s$^{-1}$,
3.80 $\pm{0.18}$ $\times$ 10$^{5}$ ergs cm$^{-2}$ s$^{-1}$). 
The bimodal structure of the quiescent CIV line profile might be due to two
different broadening mechanisms and  velocity fields at 10$^{5}$ K. 
The width of the narrow component can be due to solar type TR transient
events (random multiple velocities, 
but with a much higher filling factor), while the broad component
could arise from frequent microflaring.
The possibility
 that the broad line component comes from a ring of
sling-shot prominences  at 2 - 3 stellar radii, deserves to be further studied.

2. The transition region of AB Dor (as seen in the CIV line emission)
 is flaring  frequently in different time scales and 
magnitudes (see Figs. \ref{fig:2D} and \ref{fig:CIV}). 
The quiescent mean flux was f$_{CIV}$ = (1.19 $\pm{0.05}$) $\times$ 
10$^{-12}$
ergs cm$^{-2}$ s$^{-1}$ on the Earth. The variability over the whole observing
run is characterized by the standard deviation 1$\sigma$ = 5.4 $\times$ 
10$^{-13}$
ergs cm$^{-2}$ s$^{-1}$, while the 1$\sigma$ of the individual data points 
in the light curve (Fig. \ref{fig:CIV}) equals to 1.6 $\times$ 10$^{-13}$ 
ergs cm$^{-2}$ s$^{-1}$. This means that 
the star spends about 15 per cent of time outside the 3$\sigma$-level from the
quiescent state. It is even possible that the quiescent state itself is composed
of many overlapping microflares, as suggested by the broad line component. 
The chromospheric lines (SIII 1533.432 and 
the blend of SiI, CI and FeII lines at 1561 \AA) remained constant within
a 3$\sigma$-level.

3. The largest impulsive flare seen in the CIV 1550 emission 
at day 14.22 reached in less than a minute the 
differential emission measure of 10$^{51.2}$ cm$^{-3}$ ( = N$_{e}^2$ V between
10$^{4.85}$ K - 10$^{5.15}$ K) and returned to the 7 times lower quiescent 
level after 5 minutes (see Table 2).
During the flare peak the surface flux (on the star) reached 
the value of F$_{CIV}$ = (5.13 $\pm{0.20}$) $\times$ 10$^6$ ergs cm$^{-2}$ 
s$^{-1}$. The total energy of the flare radiated
in the CIV 1549 line was 1.65 $\pm{0.10}$ $\times$ 10$^{31}$ ergs. 
The profile of the line during this flare
(3 min average) was modelled with the quiescent profile {\it plus} a
broad (FWHM = 800 km s$^{-1}$) and blue-shifted (--190 km s$^{-1}$) Gaussian. 
During the flare,
slightly blue-shifted (--50 km s$^{-1}$) and narrow (FWHM = 55 km s$^{-1}$)
absorptions appeared, having equivalent widths of 0.3 \AA~ and
0.2 \AA~ in the 1548 and 1550 \AA~ lines, 
respectively (after the quiescent profile was 
subtracted). 

4. The impulsive transition region flare at day 14.22 can be qualitatively
understood with the help of the
    hydrodynamical models by \markcite{mari89} Mariska \ea (1989), based on 
    similar
time scales of the event and the models.
In this comparison  
the flaring loop foot point area covered (0.5 - 1.0) \%
of the total stellar surface.
The total 
number of injected electrons 
can be estimated around  10$^{40}$ needing a reservoir of a 
 volume of 10$^{30}$ cm$^{3}$ at the apex of a typical coronal loop
 (the reconnection area).   

\acknowledgments

We thank Dr David Soderblom for the assistance to study the wavelength scale
of GHRS and Drs Jeffrey L. Linsky and  Thomas R. Ayres, 
the referees, for valuable comments and 
critisism. We are grateful to Prof. P. Maltby and Dr. N. Brynildsen for making available  
Spacelab solar data.
\clearpage


\begin{table}
\dummytable\label{tbl:1}
\end{table}


\begin{table}
\dummytable\label{tbl:2}
\end{table}

\clearpage


\begin{table}
\dummytable\label{tbl:3}
\end{table}


\clearpage

\figcaption[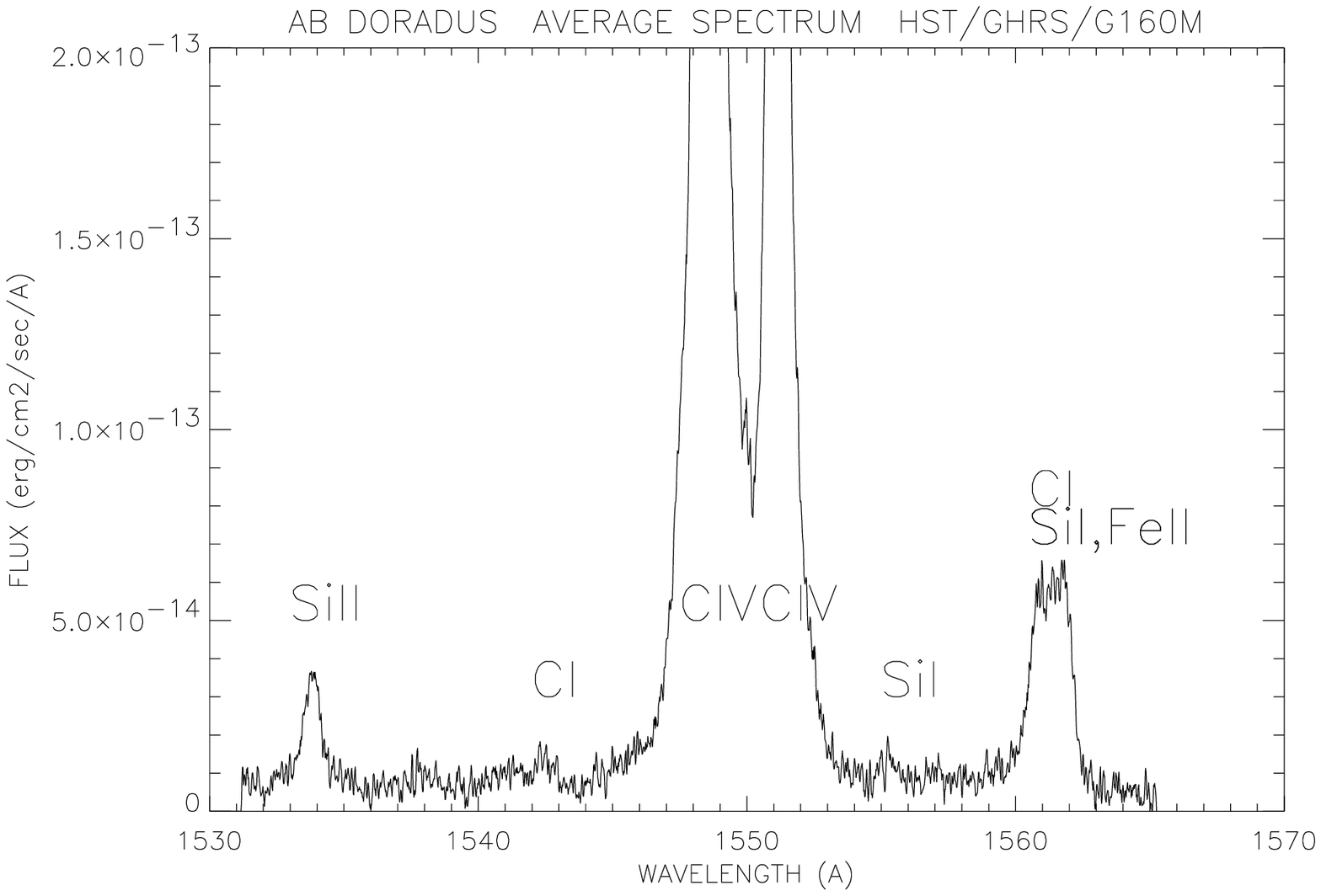]{The time-averaged spectrum of AB Doradus
taken with the GHRS of the Space Telescope using the grating G160M. \label
{fig:averspec}}

\figcaption[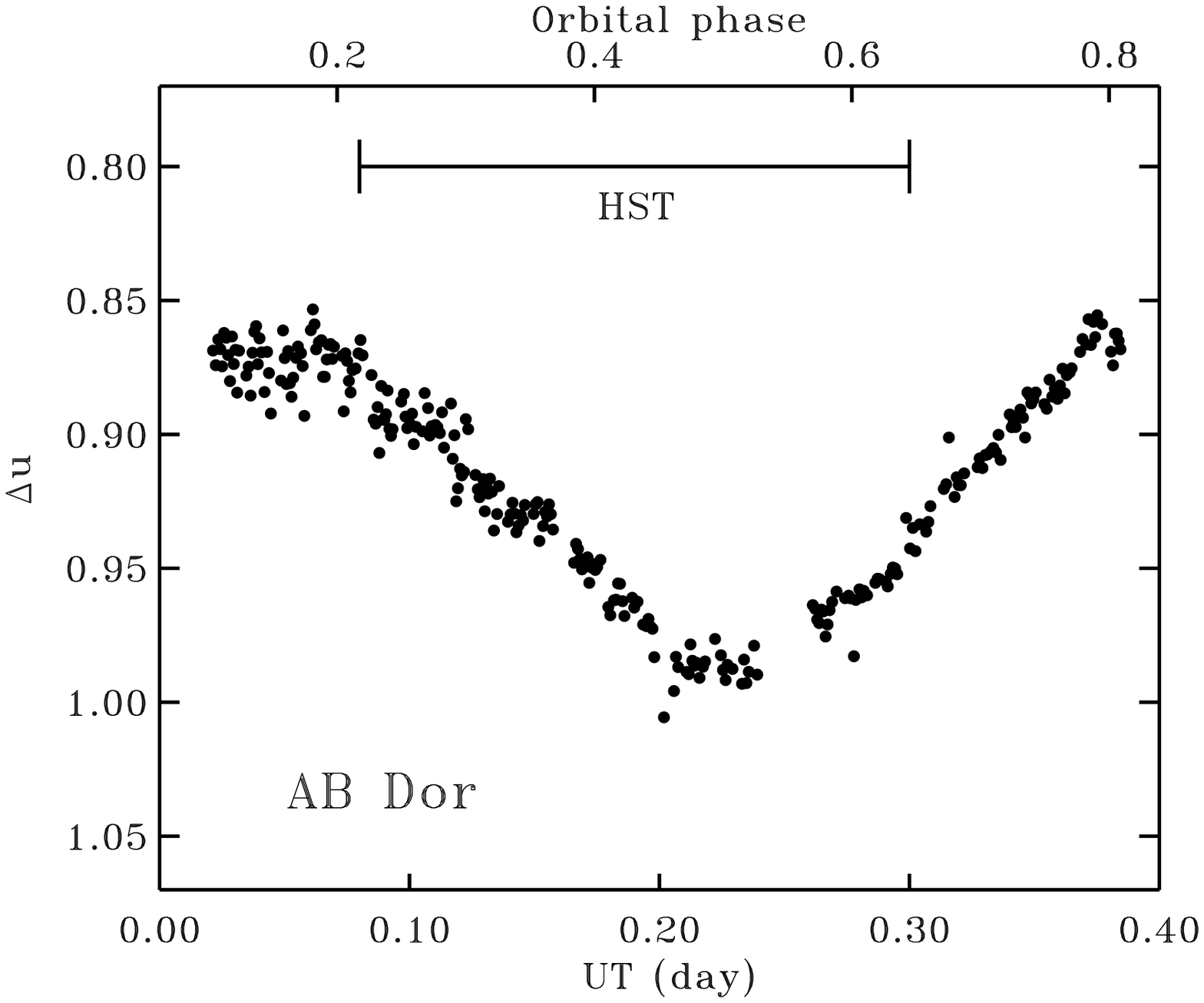]{The optical U-band light curve of AB Dor during 
the present {\it HST} observations covering the day fraction 0.08 - 0.30, 
corresponding to rotational phases 0.21 - 0.64 (Eq. \protect\ref{eq:eph}). 
A photospheric dark  spot region  seen face-on at phase 0.5 is responsible for the minimum. 
\label{fig:Ucurve}}

\figcaption[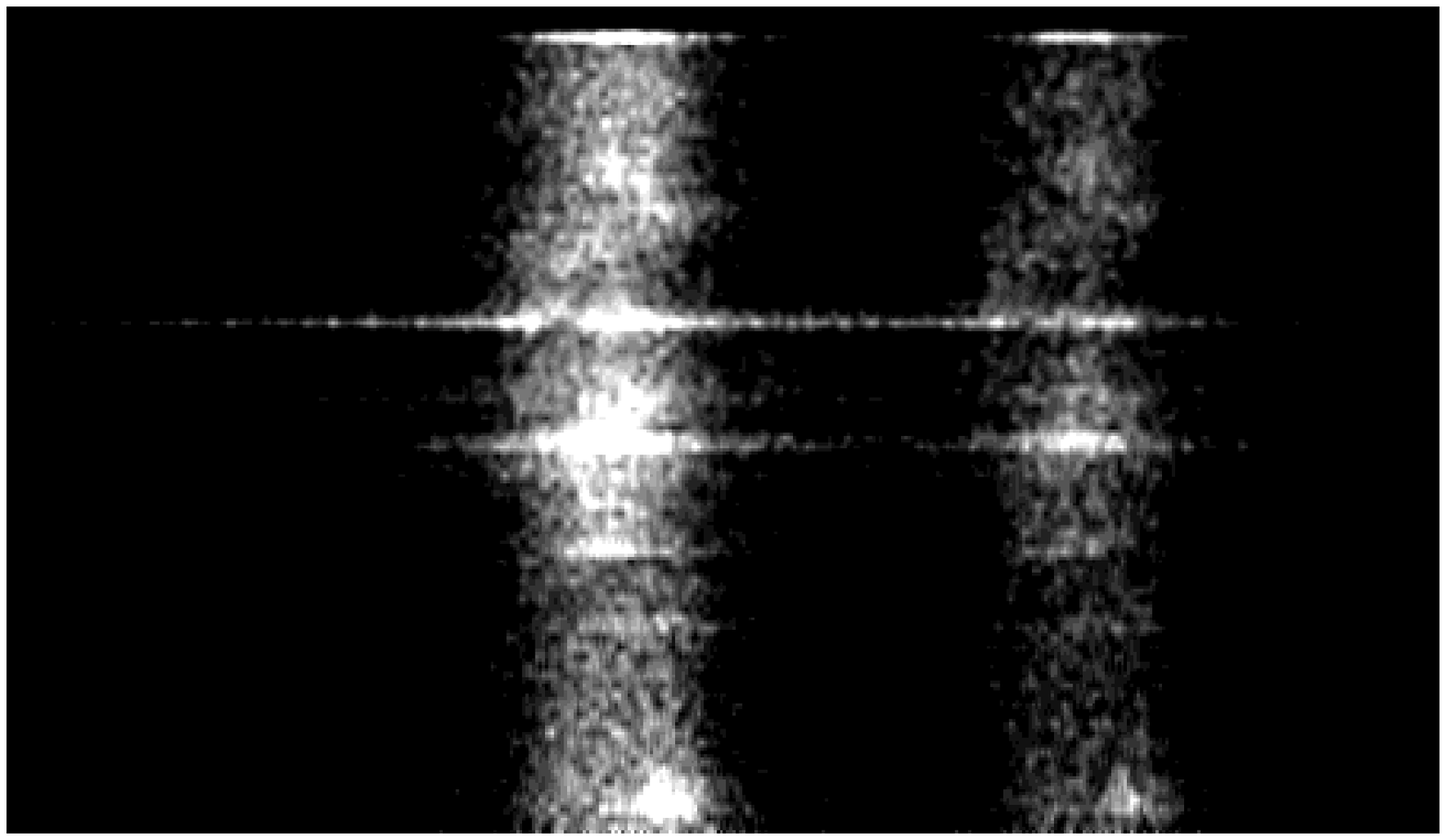]{The dynamic spectrum of AB Dor at the CIV 1549 
doublet. The time goes vertically upwards (total 5.5 hours) and the wavelength
increases from left to right. The impulsive     flare at day 14.22
discussed in Ch. 3 is seen as the brightest horizontal spike. \label{fig:2D}}


\figcaption[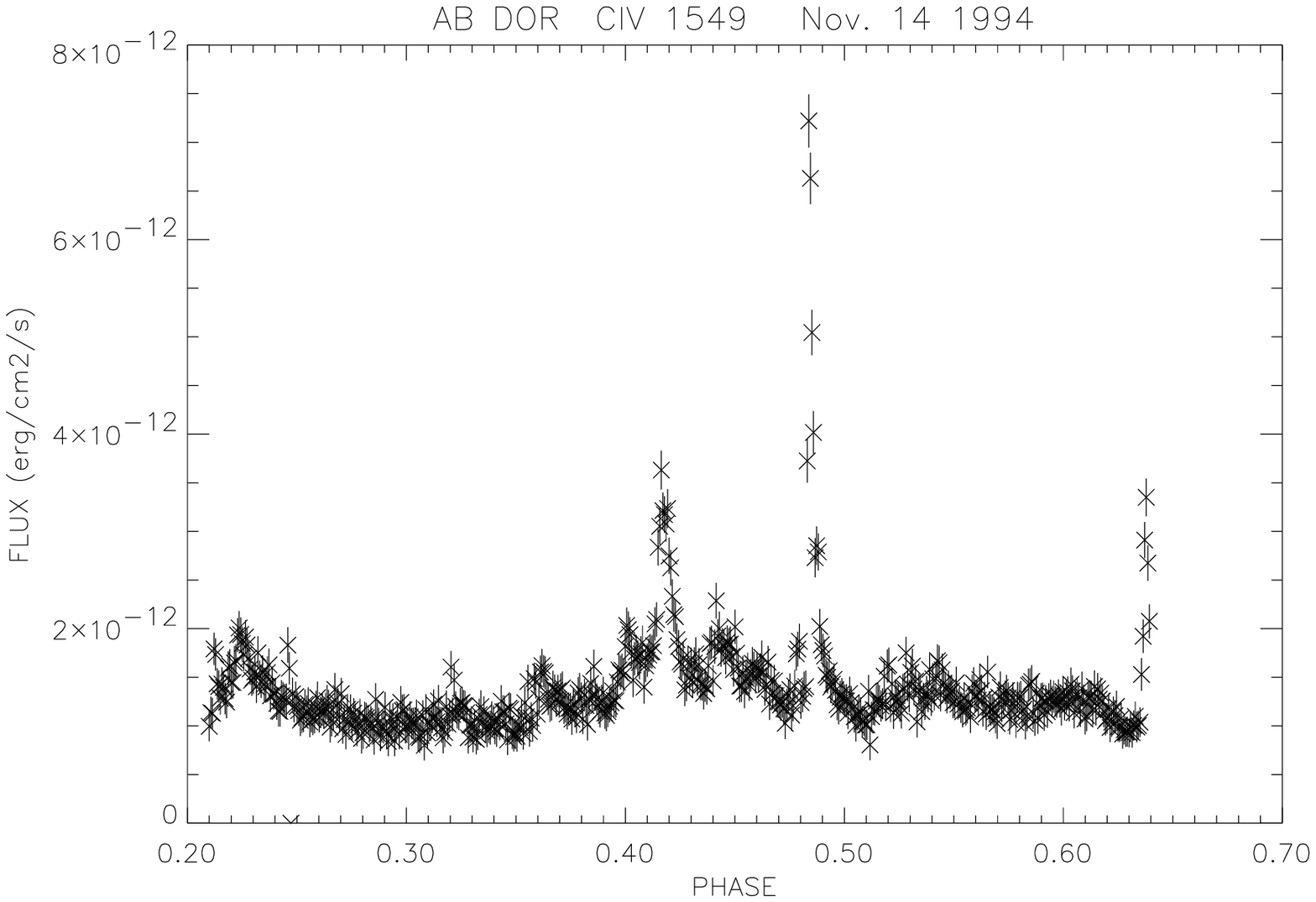]{ The light curve of the total CIV 1549 doublet 
with 27 s binning.  \label{fig:CIV}}


\figcaption[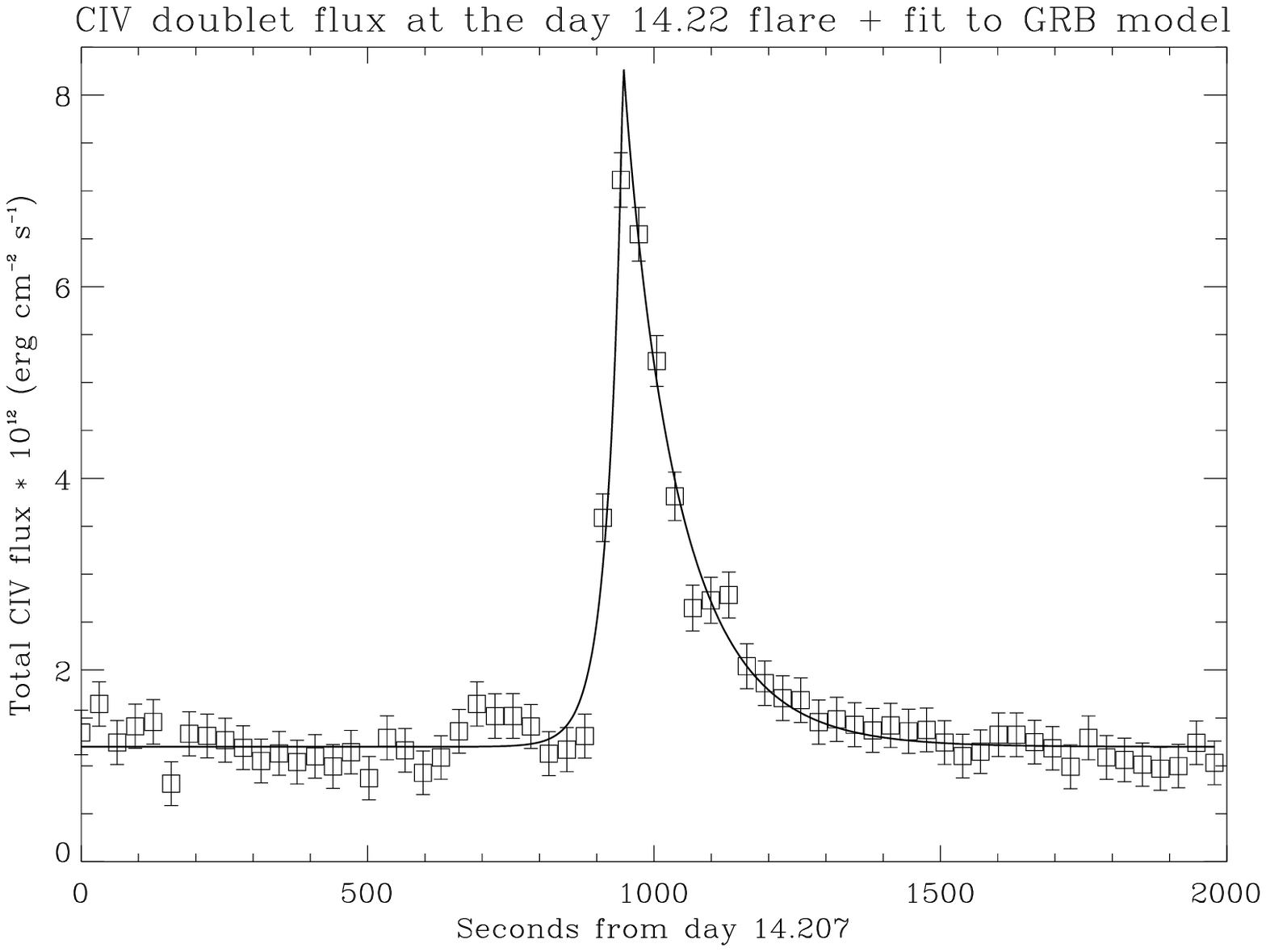]{The CIV light curve around the strongest flares. 
The model fits are included (Eq. 2).
(a) Day 14.185 flare. Seconds are counted from the  day fraction 14.173.
(b) Day 14.22 flare. Seconds are counted from the day fraction 14.207. 
\label{fig:flarefits}}

\figcaption[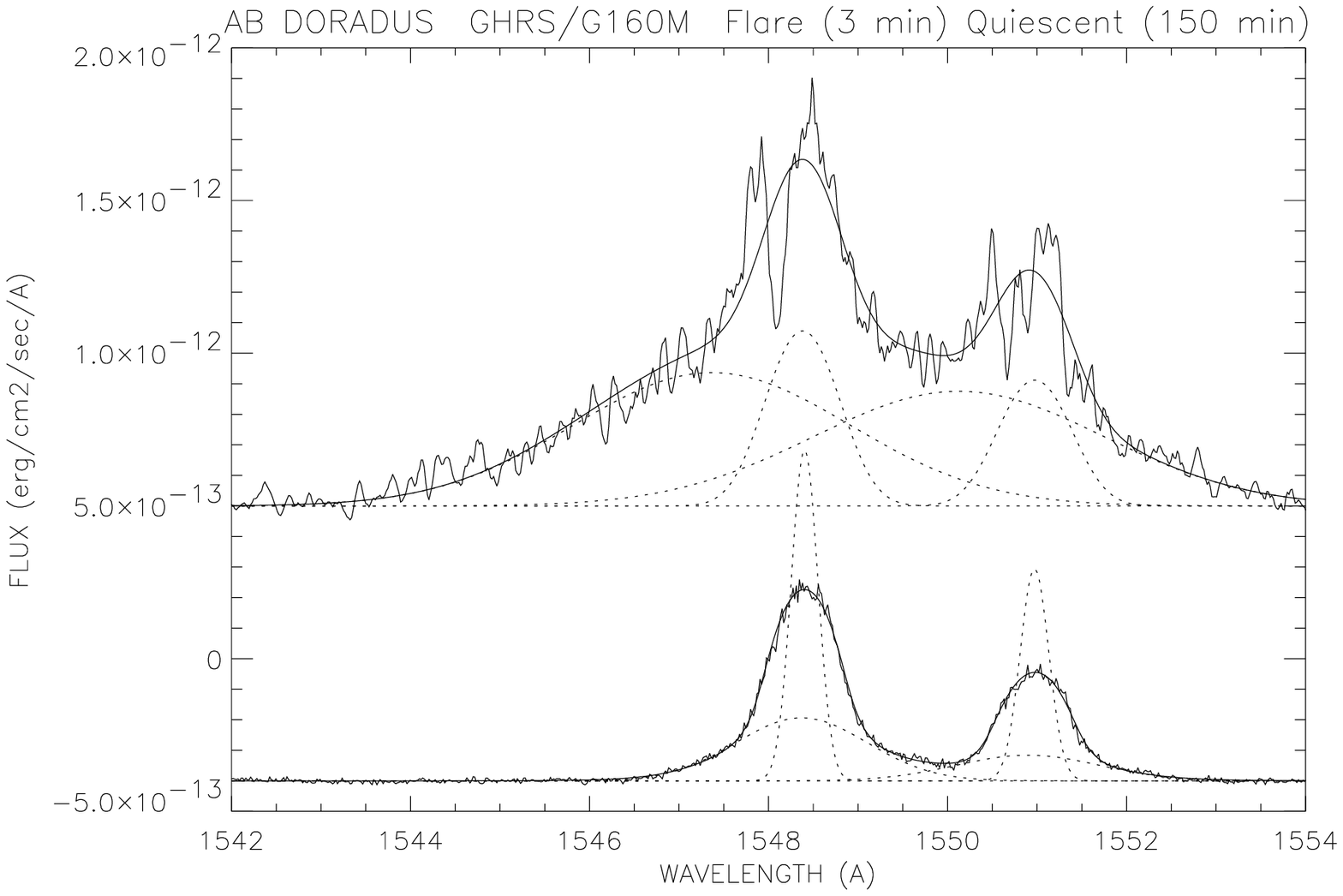]{Spectra of AB Dor at CIV 1549 during the
strongest flare and the quiescent state.
Upper: The time averaged (exp = 3 min)
flare spectrum at day 14.22 Nov 1994, with 4-Gaussian
fits (the dashed lines). 
Lower: The quiescent spectrum accumulated during 150 
minutes of observation with flaring extracted. The intrinsic (non-rotating) 
4-Gaussian components
(dashed lines) were further broadened with the 90 km s$^{-1}$ rotational and
15 km s$^{-1}$ instrumental profiles to obtain the best fit shown by the solid 
line (difficult to see against the stellar spectrum). Note that the flaring
narrow component is almost identical to the quiescent
profile. The continuum is shifted by 5.0 and 
   -4.0 (10$^{-13}$) for the flare and quiescent spectra, respectively.
\label{fig:lineprofs}}

\end{document}